\documentclass[prl,aps,10pt,twocolumn]{revtex4-1}

\usepackage{latexsym}
\usepackage{amstext}
\usepackage{amsmath}
\usepackage[normalem]{ulem}
\usepackage{subfigure}
\usepackage{epsfig, graphicx}
\usepackage{color}
\usepackage{amsfonts,bm,color}
\usepackage{verbatim}
\usepackage{float}
\usepackage{latexsym}
\usepackage[svgnames]{xcolor}
\usepackage{colortbl}
\usepackage{float}
\usepackage{wrapfig}
\usepackage{multirow}

\newcommand{\ds}{\displaystyle}
\newcommand{\beq}{\begin{eqnarray}}
\newcommand{\eeq}{\end{eqnarray}}
\newcommand{\beqq}{\begin{eqnarray*}}
\newcommand{\eeqq}{\end{eqnarray*}}
\newcommand{\p}{\partial}

\newcommand{\eps}{\varepsilon}
\newcommand{\x}{\mbox{\boldmath$x$}}

\newcommand{\fF}{\mathcal{F}}

\begin{document}
\title{Electrical transient laws in neuronal microdomains based on electro-diffusion}
\author{J. Cartailler and D. Holcman}
\affiliation{Ecole Normale Sup\'erieure,75005 Paris, France.}
\date{\today}
\begin{abstract}
The current-voltage (I-V) conversion characterizes the physiology of cellular microdomains and reflects cellular communication, excitability, and electrical transduction. Yet deriving such I-V laws remains a major challenge in most cellular microdomains due to their small sizes and the difficulty of accessing voltage with a high nanometer precision. We present here novel analytical relations derived for different numbers of ionic species inside a neuronal micro/nano-domains, such as dendritic spines.  When a steady-state current is injected, we find a large deviation from the classical Ohm's law, showing that the spine neck resistance is insuficent to characterize electrical properties. For a constricted spine neck, modeled by a hyperboloid, we obtain a new I-V law that illustrates the consequences of narrow passages on electrical conduction. Finally, during a fast current transient, the local voltage is modulated by the distance between activated voltage-gated channels. To conclude, electro-diffusion laws can now be used to interpret voltage distribution in neuronal microdomains.
\end{abstract}	
\maketitle
Electro-diffusion in cellular nanodomains has recently regained interest due to the increase in fluorescent voltage dye indicators precision \cite{Cohen}, development of nanopipettes \cite{Krisna} and in parallel the extension of the associated Poisson-Nernst-Planck theory to compute voltage. This theory was previously succesful to study ionic fluxes and gating of voltage-channels \cite{Bezanilla}, because at the nanometer scale, cylindrical symmetry of a channel reduces computations to a one-dimensional model for the electric field and charge densities in the channel pore \cite{Eisenberg,EKS}. However, cellular microdomains involve in general two- and three-dimensional neuronal geometries \cite{Peskin,Sejnowski}, which make the analysis of the PNP equations much more complicated than in the cylindrical geometry of a channel pore \cite{Qian}. Nevertheless, the PNP theory is now used to study the current-voltage relation in several cellular compartments such as the synaptic cleft \cite{Sylantyev,Savtchenko2,Sylantyev2}, dendritic spines \cite{cartailler2018deconvolution} and many others \cite{HY2015}. Indeed, the classical cable theory or simply the electrical resistance are insufficient to describe the electrical properties of a dendritic spine, which can be seen as a cellular micro-electrolyte. In general computing the I-V relation has remained controversial especially about the order of magnitude of an effective resistance, ranging from few to thousands of Mega Ohms \cite{Popovic,Loew,Kwon,Beaulieu}. Computing the I-V relation in the context of cellular physiology is relevant because any long lasting changes are a signature of a form of plastic changes, and having a precise relation would help clarifying synaptic plasticity, that underlies learning and memory.\\
In the absence of local electro-diffusion, we previously showed that geometrical features, such as curvature or narrow funnels can modulate the electrostatic properties of non-electroneurotral electrolytes \cite{cartailler2017analysis} and could even generate local potential differences along the surface of a corrugated cylinder, between neighbording points of positive and negative curvature \cite{cartailler2017electrostatics}.\\
We report here novel I-V relations when a current is injected for various geometrical structures such as a dendritic spine modeled as a ball connected to a narrow neck. A constricted neck is modeled by a hyperboloid containing a narrow passage.  The new relation we derived show significant deviation compared to the classical Ohm's law. During a fast current transient, we further investigate how the local voltage is modulated and how it depends on the distance between the activated neighbording voltage-gated channels.\\
{\noindent \bf Poisson-Nernst-Planck (PNP) theory for electro-diffusion model.}
The Poisson-Nernst-Planck (PNP) theory describe the coarse-grained dynamics of charged ions, accounting for the coupling between the ionic flow and electrostatic forces \cite{EKS,Nadler,Nitzan,Orland,STINCHCOMBE}. For an electrolyte composed of two monovalent ions, the voltage $V$ and the concentration of positive $c_p$ and negative $c_m$ ions are described by{\small
\beq\label{3D_EQ_PNP}
\Delta V(\x,t)&=&-\frac{\fF}{\eps\eps_0}\left( c_p(\x,t) - c_m(\x,t) \right)\\
\frac{\p c_p}{\p t}(\x,t) &=&D_p\nabla\cdot \left(\nabla c_p(\x,t) +\frac{e}{k_B T}c_p(\x,t) \nabla V(\x,t) \right)\nonumber\\
\frac{\p c_m}{\p t}(\x,t) &=&D_m\nabla \cdot\left(\nabla c_p(\x,t) -\frac{e}{k_B T}c_m(\x,t) \nabla V(\x,t) \right),\nonumber
\eeq}
where $\fF$ is the Faraday's constant, $\eps\eps_0$ the electrolyte permitivity, $k_BT/e$ the thermal voltage and $D_p$ and $D_m$ are the diffusion coefficient for positive and negative charge respectively (see table \ref{table2} below). The present model can be extended to two cations K$^+$, Na$^+$ and one generic anions A$^-$.\\
\begin{figure}[http!]
	\center
	\includegraphics[scale=0.2]{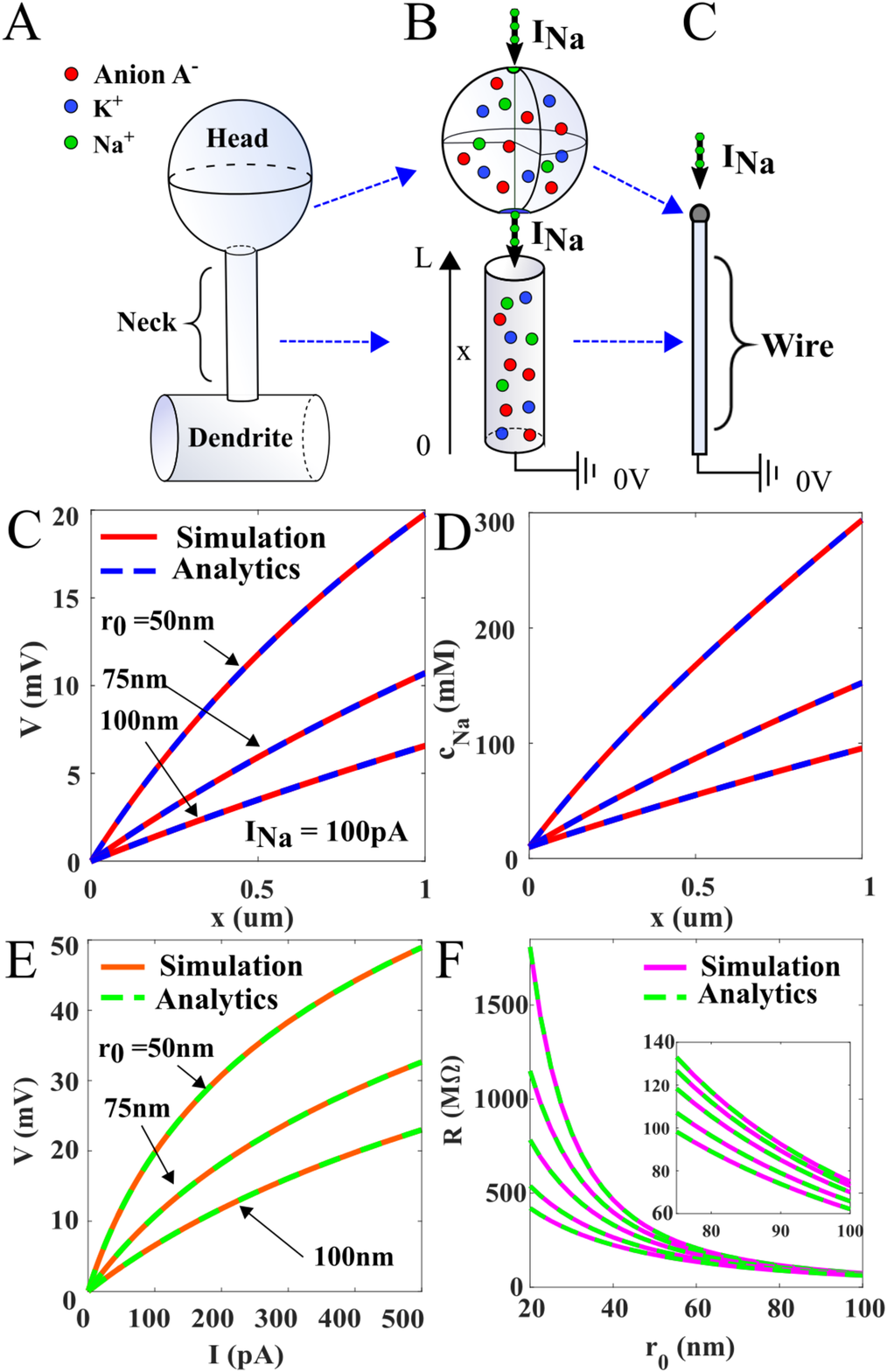}
	\caption{{\small {\bf Steady-state voltage and charge distribution in a three ionic species cylindrical electrolyte}
			{\bf A.} Spine geometry containing an electrolyte composed of Na$^+$, K$^+$ and a monovalent anions A$^-$ in a cylinder of length $L$ and radius $r_0$. The injected current $I$ at the top of the cylinder is composed exclusively of Na$^+$ ions. {\bf C-D.} Voltage and concentration distributions, computed numerically (red) and analytically from eq. \eqref{Voltage_Sym2} (dashed blue) along the $x-$axis for $I_{Na}=\{100 \}pA$ and $r_0=50/75/100nm$ ({\bf B}).
			{\bf E-F.} I-V relation and resistance properties computed numerically (red) \eqref{3D_EQ_PNP} and analytically(green) from \eqref{Resistance1}.
			\label{fig1}}}
\end{figure}
{\noindent \bf Voltage-Current relation for a dendritic spine.}
To find the I-V relation for a steady-state injected current $I$ made of positive charges, we use the reduction of the spine head geometry to a single point \cite{cartailler2018deconvolution} (Fig. \ref{fig1}A-B), because the voltage is constant (except in a boundary layer of the region where the current is injected) . We further approximate the narrow homogeneous cylindrical neck to a segment of length $L$. To derive the I-V relation, we reduce equations \ref{3D_EQ_PNP}, to a one dimension system with two species. The steady-state equations are{\small
\beq\label{PNP_1D_STD}
\frac{d^2 c_p(x)}{d x^2} &=& -\frac{d}{d x}\left( c_p(x)\frac{d u(x)}{d x} \right) \\
\frac{d^2 c_m(x)}{d x^2} &=& \frac{d}{d x}\left( c_m(x)\frac{d u(x)}{d x} \right) \nonumber
\eeq}
where $u(x)=\frac{eV(x)}{kT}$ is the non-dimensionalized voltage. The boundary conditions are{\small
\beq\label{BVP}
\left. \frac{d c_p(x)}{d x}+c_p(x)\frac{d u(x)}{d x}\right|_{x=L} &=&\frac{I}{\pi r_0^2 D_p F} \\
\left. \frac{d c_m(x)}{d x}-c_m(x)\frac{d u(x)}{d x}\right| _{x=L}&=&0 \nonumber\\
c_p(0) = c_m(0) =C,\nonumber
\eeq}
where the first term corresponds to the injected current at the head, $r_0$ is the radius of the neck and $C$ is the fixed concentration imposed for positive and negative species at the dendrite. A direct integration of eq. \eqref{PNP_1D_STD} using \eqref{BVP}, leads to the Boltzmann distribution for negative charges $c_m(x)=C e^{\ds u(x)}$ in the one-dimensional segment. At this stage, we assume electro-neutrality at all spatial scale {\small
\beq\label{Electro_neutral}
c_p(x)&=&c_m(x).
\eeq}
Using \eqref{Electro_neutral} in \eqref{PNP_1D_STD}, we obtain
$ c_p(x)=C  +\frac{Ix}{2 S D_p F}$ and the voltage {\small
\beq\label{Voltage_Sym}
V(x)&=&\frac{kT}{e}\ln \left(1 +\frac{Ix}{2 C \pi r_0^2 D_p F} \right).
\eeq}
We conclude that the electrical resistance from eq. \eqref{Voltage_Sym} is
{\small \beq\label{Resistance1}
R(I)&=\ds\frac{|V(L)-V(0)|}{I}&=\frac{kT}{I e}\ln \left(1 +\frac{IL}{2 C  \pi r_0^2 D_p F} \right).
\eeq}
In the limit $\frac{I}{S}\ll1$, the leading order term in \eqref{Resistance1} reduces to Ohm's law:{\small
\beq\label{Ohm_law}
R(I)&=& \frac{kT }{eF}\frac{ L}{2 \pi r_0^2 C  D_p }+O\left(\frac{I}{S}\right).
\eeq}
We validated expression \ref{Ohm_law} with repect to three-dimensional simulations of equation \ref{3D_EQ_PNP} (Fig. \ref{fig1}C-D and E-F). To conclude, Ohm's law for electrolyte is valid only if the injected current $I$ is small or the section surface $S$ is large. In practice $O\left(\frac{I}{S}\right)=O(1)$ and thus the Ohm's approximation \eqref{Ohm_law} is not applicable and should be replaced by \eqref{Resistance1}. \\
{\noindent \bf Voltage distribution with two positive specie Na$^+$, K$^+$.}
Equations \ref{PNP_1D_STD} can be used with three species K$^+$, Na$^+$ and anions A$^-$, when a steady current composed of sodium ions is injected at one end, leading to {\small
\beq\label{NP_c1}
 c_{Na}''(x)  &=& -\left[ c_{Na}(x)u'(x)  \right]_{x}\\
 c_{K}''(x)  &=& -\left[ c_{K}(x)u'(x)  \right]_{x} \nonumber\\
 c_m''(x)  &=& \left[ c_m(x)u'(x)  \right]_{x}, \nonumber
\eeq}
with the boundary conditions:
{\small\beq\label{3ions_bvp}
\left. \frac{d c_{Na}(x)}{d x}+c_{Na}(x)\frac{d u(x)}{d x}\right|_{x=L} =&\frac{I_{Na}}{\pi r_0^2 D_{Na}  F}\\
\left. \frac{d c_{K}(x)}{d x}+c_{K}(x)\frac{d u(x)}{d x}\right|_{x=L} =&0 \nonumber\\
\left. \frac{d c_{A}(x)}{d x}-c_{A}(x)\frac{d u(x)}{d x}\right| _{x=L}=&0 \nonumber \\
c_{Na}(0) = C_{Na}\mbox{,} c_{K}(0) =&\, C_{K} \mbox{,}  c_{A}(0) = C_{A},\nonumber
\eeq
}with $C_{A}=C_{Na}+C_{K}$. The no-flux boundary condition \eqref{3ions_bvp} implies that the concentrations $c_{K}$ and $c_{A}$ followed are Boltzmann distributions. Assuming the local electroneutrality at all spatial scale, we have {\small
\beq\label{positive_sum}
c_{Na}(x)&=&C_{A}e^{u(x)}-C_{K}e^{-u(x)}.
\eeq}
Using \eqref{positive_sum}, we obtain an expression for the normalized potential {\small
\beq\label{non_dim_volt_2}
u(x) &=&\ln \left( \frac{c_{Na}(x)+\sqrt{c_{Na}^2(x) +4C_{K}C_{A}}}{2C_{A}} \right),
\eeq}
and the flux boundary condition in \eqref{3ions_bvp} leads to the expression for the concentration of $Na^+$:
{\small \beq\label{Na_conc}
c_{Na}(x)  &=&C_{A} +\frac{Ix}{2 \pi r_0^2 D_{Na} F} -\frac{2C_{K}C_{A}}{2C_{A} +\frac{Ix}{\pi r_0^2 D_{Na} F}}.
\eeq
}
The last descreasing term in expression \eqref{Na_conc} is an expression of the coupling between the potassium and the anionic concentration, when the injected current is due to the sodium ions only. Indeed, potassium ions are expeled due to the positive sodium ions.  Surprizingly, the voltage is not affected by this coupling, indeed using $\tilde c(x) = c_{Na}(x)+c_{K}(x)$, we reduced the system of equations \eqref{3ions_bvp} to two species and using  \eqref{Voltage_Sym}, we get {\small
\beq\label{Voltage_Sym2}
V(x)&=&\frac{kT}{e}\ln \left(1 +\frac{Ix}{2 C_{A} \pi r_0^2 D_{Na} F} \right),
\eeq}
and the effective resistance is given by 
{\small \beq\label{Resistance2}
R(I)&=&\frac{kT}{e}\ln \left(1 +\frac{IL}{2 C_{A} \pi r_0^2 D_{Na} F} \right).
\eeq}
To conclude, equations \eqref{Voltage_Sym2} and \eqref{Resistance2} show that several monovalent ionic species do not affect neither the voltage nor the resistance of an electrolyte.  However, the type of the injected ions influence the voltage distribution only through the diffusion coefficient of the injected specie (here $D_{Na}$).\\
\begin{figure}[http!]
	\center
	\includegraphics[scale=0.37]{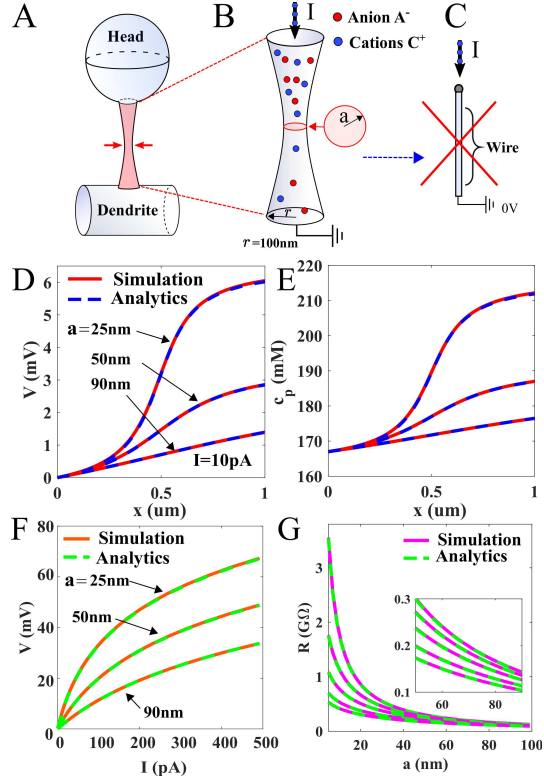}
	\caption{{\small {\bf Voltage and concentration distribution in a domain with a constricted section}
			{\bf A-B-C.} Constricted domain formed by a hyperboloid \eqref{hyperboloide}, where $a$ is radius of the constriction. It is not possible to reduce the geometry to a segment.
			{\bf D,E.} Voltage and positive charge distribution computed numerically with an injected current $I_p=50pA$ along the symmetry axis obtained numerically (red) and with \eqref{Volt_hyper} (dashed blue) for several radii $a=\{25, 50, 90 \}nm$, $I_p=10pA$.
			{\bf F,G.} I-V relation and effective Resistance computed numerically (red) along the symmetry axis and analytically from eq. \eqref{concentration_hyper}. Paremeters are similar to {\bf D,E}.
\label{fig2}}}
\end{figure}
\begin{figure*}[http!]
	\center
	\includegraphics[scale=0.50]{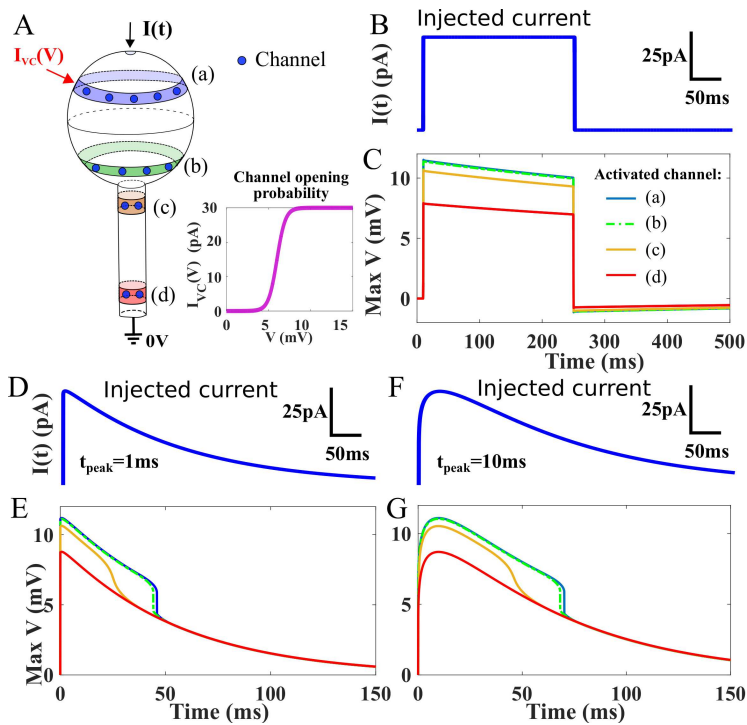}
	\caption{{\small {\bf Voltage response to a transient injected current $I(t)$ in a spine shaped domain, containing voltage-gated channels on the surface}
	{\bf A.} The injected current $I(t)$ is at the top of the spine, while the voltage is grounded ($0V$) at the base. The voltage channels are positive in narrow (colored) bands, leading to a transient current $I_{VC}(V)$, where $V$ is the potential averaged in the synmmetrical band. Inset :$I_{VC}(V)$ current activation curve.
	{\bf B-C.} Injected current $I(t)$ is a step-function, starting at $10ms$.Voltage difference $\Delta V(t)= \max_{\x\in\Omega} V(t)$ computed when a current $I_{VC}(V)$ is injected at the head top (blue) and bottom (green), the top (orange) and the bottom (red) of the neck. {\bf D-E/F-G} Transient injected current $I(t)=I_0 \,t^\alpha \exp{\left(-\frac{\alpha t}{t_{peak}}\right)}$ where $\alpha = 1$, $t_{peak}=10ms$ and $I_0$ is calibrated such as  $I(t_{peak})=50pA$.
\label{fig3}}}
\end{figure*}
{\noindent \bf Steady-State PNP solutions in constricted neck.}
To account for a possible physical constriction in the neck of a spine or any other neuronal microdomains, containing positive and negative ions, we model this geometry as a hyperboloid of length $2L$ with an elliptic base
\beq\label{hyperboloide}
\frac{y^2}{a^2}+\frac{x^2}{b^2}&\leq& \frac{z^2}{c^2}+1, \, z\in[-L; L].
\eeq
(Fig. \ref{fig2}A), where the constriction has a surface $\pi ab$. Following our previous anlaysis in one dimension \ref{PNP_1D_STD}, because the concentration of negative species $c_n$ depends only on the $z-$variable, assuming electroneutrality, we obtain that $c_p$ will also depends only on z. Thus the flux boundary condition \ref{BVP} lead to,
{\small \beq\label{BVP2}
\left. \frac{d c_p(z)}{d z}+c_p(z)\frac{d u(z)}{d z}\right|_{z=L} &=&\frac{I}{\pi ab(1+\frac{L^2}{c^2}) D_p F} \\
\left. \frac{d c_m(z)}{d z}-c_m(z)\frac{d u(x)}{d x}\right| _{z=L}&=&0 \nonumber\\
c_p(-L) = c_m(-L) =C,\nonumber
\eeq}
and we obtain {\small
\beq\label{diff_hyper}
\frac{dc_p(z)}{dz}&=&\frac{I}{2 \pi ab(1+\frac{z^2}{c^2}) D_p F}\\
c_p(-L)&=&C_p.\nonumber
\eeq}
leading to
{\small
\beqq\label{concentration_hyper}
c_p(z)&=&\frac{Ic}{2 \pi D_p F ab}\left(\arctan \frac{L}{c} + \arctan \frac{z}{c}  \right)  +C_p.
\eeqq
}
Thus the voltage followed from the Boltzmann distribution \eqref{concentration_hyper},
{\small
\beqq\label{Volt_hyper}
V(x)&=&\frac{k T}{e}\ln \left(1+ \frac{Ic}{2C_p \pi D_p F ab}\left(\arctan \frac{L}{c} + \arctan \frac{z}{c}  \right) \right).
\eeqq
}
We conclude that the I-V relation is characterized by the current-dependent resistance {\small
{\small
\beq\label{Res_hyper}
R_h(I)=\ds\frac{k T}{Ie}\ln \left(1+ \frac{Ic}{ C_p \pi D_p F ab} \arctan \frac{L}{c} \right).
\eeq}
We confirm this new relation using numerical simulation (in dimension 3) in Fig. \ref{fig2}. To conclude, a single local constriction can increase drastically the effective resistance, which could be mediated by organelle located in the spine neck such as vesicle ot spine apparatus \cite{korkotian2011synaptopodin}.\\
{\noindent \bf Transient properties.}
We now study the properties when a transient step function for the current is generated in a cap around the north pole (with a radius $r=10nm$) of a dendritic spine. We solve numerically equations \ref{3D_EQ_PNP}  (Fig. \ref{fig3}A). Except the cap, the rest of the boundary is reflecting for the potential and the ionic fluxes. At the dendrite, the concentration is maintained fixed. Interestingly, for a current with an ampltideu 50pA, we obtain a transient voltage with a maximum of 6mV (inside the spine). However, this transient is modified when voltage-gated channels organized in rings, can be activated (the activation curve is shown in Fig. \ref{fig3}A inset) when there are located inside the head. When channels are located at the entrance of the neck, they are activated with a delay of few hundreds of ms, but not at the base of the neck (fig. \ref{fig3}B-G). We conclude that voltage-gated channels located in the spine head can increase transiently the voltage amplitude generated by a synaptic current.\\
{\noindent \bf Conclusion.}
To conclude, under the eletro-neutrality assumption at all scale, when a steady-state current is injected, we derived here novel I-V relation in a dendritic spine, which is a key microdomain underlying neuronal communication \cite{Yuste}. We further show how the I-V relation is affected by a neck constriction, which is often the case due to the presence of organelles such as a spine apparatus \cite{korkotian2011synaptopodin}. \\
Injected a transient current revealed that the changes in the voltage can be sufficient to activate voltage-gated channels inside the spine head, but not located at the end of the spine neck. The main consequences of obtaining these laws is to show that local changes in the nano/microdomain geometry can modify the I-V relation, a part of nanophysiology that has often been neglected \cite{Rusakov3}, and thus contribute to the modulation of synpatic plasticity.
	\begin{table}[!http]
		\caption{Simulation parameters.}
		\begin{center}\label{table2}
			\scalebox{0.8}{
				\begin{tabular}{ l l l}
					\hline  Parameter & \, \, \, \, Description & Value\\
					$z$& Valence of ions &\mbox{z=1} \mbox{(for Na$^+$ and K$^+$)}\\
					$D_p$&  Diff.  coeff. for $+$ charges & $D_p=D$\\
					$D_m$&  Diff.  coeff. for $-$ charges & $D_m=D$\\
					$\lambda$ & Tortuosity & \\
					$C_p$&  $+$ charge concentration  & $C_p=167 $mol/m$^3$ \cite{Hille} \\
					$C_m$&  $-$ charge concentration  & $C_m=167 $mol/m$^3$\cite{Hille}\\
		            $C_{Na}$&  $+$ charge concentration  & $C_p=12 $mol/m$^3$\cite{Hille} \\
	                $C_{K}$&  $+$ charge concentration  & $C_p=155 $mol/m$^3$ \cite{Hille} \\				
					$kT/e$&  Thermal voltage  &   $25.26$V\\
					$\eps$&  Dielectric constant  &   $\eps=80$ (water)\\
					$\eps_0$&  Abs. Dielectric constant  &   $\eps_0=8.8 \cdot10^{-12}$ F/m\\
					$\mathcal{F}$ &Faraday's constant & $ \mathcal{F}=96485C/mol$\\
					\hline
				\end{tabular}
			}
		\end{center}
\end{table}

%
\bibliographystyle{apsrev4-1}
\bibliography{Biblio_transient}
\end{document}